\documentclass[11pt]{article}
\usepackage{cite}
\usepackage{amsmath,amsfonts,amssymb,graphicx,calrsfs}
\usepackage[small,bf,hang]{caption}
\usepackage{slashed}
\usepackage{mathabx,amsmath}
\usepackage{color}

\usepackage[vcentermath]{youngtab}
\usepackage{mathrsfs}

\usepackage{dsfont}
\usepackage[Symbol]{upgreek}

\def\hybrid{
        \topmargin -20pt
        \oddsidemargin 0pt
        \headheight 0pt \headsep 0pt
        \textwidth 6.25in 
        \textheight 9.5in 
        \marginparwidth .875in
        \parskip 5pt plus 1pt \jot = 1.5ex}

\hybrid

\usepackage{color}
\definecolor{red}{rgb}{1,0,0}
\definecolor{lred}{rgb}{0.3,0,0}
\definecolor{green}{rgb}{0,0.6,0}
\definecolor{blue}{rgb}{0,0,1}
\definecolor{violet}{rgb}{0.8,0,0.8}

\usepackage[usenames,dvipsnames]{xcolor}

\usepackage[colorlinks=true,
pdfstartview=FitV,
linkcolor= BrickRed,
citecolor= BrickRed,
urlcolor= BrickRed,
hyperindex=true,
hyperfigures=true]
{hyperref}
\hypersetup{linktoc=page}

 \csname
@addtoreset\endcsname{equation}{section}

\def\moth{\mathsurround=0pt}
\newdimen\zo \zo=0pt

\def\tick{\leaders\hrule height 0.5ex depth 0pt \hskip 0.5pt}
\def\upboxfill{$\moth \setbox\zo\hbox{\tick}%
  \hskip 3pt\hbox to 0pt{$\tick$\hss}\hrulefill \hbox to 7.5pt{$\tick$\hss}$}

\def\dtick{\leaders\hrule height .34pt depth 0.5ex \hskip 0.5pt}
\def\downboxfill{$\moth \setbox\zo\hbox{\dtick}%
  \hskip 2pt\hbox to 0pt{$\dtick$\hss}\hrulefill \hbox to 2pt{$\dtick$\hss}$}

\def\be#1\ee{\begin{align}#1\end{align}}

\def\bec{\begin{center}}
\def\ec{\end{center}}

\def\bea{\begin{eqnarray}}
\def\eea{\end{eqnarray}}
\def\ba{\begin{array}}
\def\ea{\end{array}}

\def\ft#1#2{{\textstyle{{\scriptstyle #1}
\over {\scriptstyle #2}}}}

\begin{document}

\begin{titlepage}
\rightline{}
\begin{center}
\vskip 1.5cm
 {\LARGE \bf{ A note on the third way consistent deformation\\[1ex]
  of Yang-Mills theory}}
\vskip 1.7cm

{\large\bf {Nihat Sadik Deger$^{a,b,c}$ and Henning Samtleben$^{d,e}$}}
\vskip .8cm

{\it  $^a$ Department of Mathematics, Bogazici University,\\
Bebek, 34342, Istanbul, Turkey}\\

\vskip .3cm
{\it $^b$ Feza Gursey Center for Physics and Mathematics,\\ 
Bogazici University, Kandilli, 34684, Istanbul, Turkey}\\
\vskip .3cm

{\it $^c$ Erwin Schr\"odinger International Institute for Mathematics and Physics,\\ 
University of Vienna, Boltzmanngasse 9, 1090, Vienna, Austria}\\
\vskip .3cm

{\it  $^d$ ENSL, CNRS, Laboratoire de physique, F-69342 Lyon, France} \\
\vskip .3cm

{\it  $^e$ Institut Universitaire de France (IUF)}
\vskip .3cm

\vskip .2cm

\end{center}

\bigskip\bigskip
\begin{center} 
\textbf{Abstract}

\end{center} 
\begin{quote}  
Three-dimensional Yang-Mills theory allows for a deformation quadratic in the field strengths
which can not be integrated to a local action without auxiliary fields. 
Yet, its covariant divergence consistently vanishes after iterating the equation, 
realizing a spin-1 analogue of `minimal massive gravity', which has been dubbed `third way consistent'.
In this note, we show that after dualization of the three-dimensional gauge fields, the model possesses
a natural action as a Chern-Simons coupled gauged sigma model. In this dual formulation, coupling to matter
and to gravity becomes straightforward. As a direct application, we derive the coupling of the model 
to ${\cal N}=1$ supergravity.

\end{quote} 
\vfill
\setcounter{footnote}{0}
\end{titlepage}

\section{Introduction}

In~\cite{Arvanitakis:2015oga} a three-dimensional gauge theory defined by the following field equations
\bea
\varepsilon^{\mu\nu\rho}\,{\cal D}_{\mu} {\tilde F}_{\nu}{}^A 
+\mu\,\tilde{F}^{\rho\,A}
&=&
-\frac1{2m}\,  \varepsilon^{\mu\nu\rho}\, f_{BC}{}^A {\tilde F}_{\mu}{}^B  \,{\tilde F}_{\nu}{}^C  
\;,
\label{YM3}
\eea
with non-abelian Yang-Mills field strength 
\bea
\tilde{F}_\mu{}^A = \frac12\,\varepsilon_{\mu\nu\rho}\,F^{\nu\rho\,A}
= \frac12\,\varepsilon_{\mu\nu\rho}\left(2\,\partial^\nu A^{\rho A}
+f_{BC}{}^A\,A^{\nu B} A^{\rho C} \right)
\;,
\eea
was considered. Here, the $f_{BC}{}^A$ are structure constants of a non-abelian gauge group $G$, and $m$ and $\mu$
are constants\footnote{Equation \eqref{YM3} with some particular choices of these constants appeared earlier in \cite{Mukhi:2011jp, Nilsson:2013fya}.}.
This is a deformation of (topologically massive) Yang-Mills theory (TMYM) as 
the l.h.s.\ of (\ref{YM3}) can be derived from its Lagrangian
\bea
{\cal L}_{\rm  TMYM} &=&
-\frac14\,{\rm tr} \left[F^{\mu\nu}  F_{\mu\nu} \right]
+\frac{\mu}{4}\,\varepsilon^{\mu\nu\rho}\, {\rm tr}\left[F_{\mu\nu} A_\rho \right]
\;.
\label{YMCS}
\eea
However, in presence of the r.h.s., the equations (\ref{YM3}) can no longer be integrated to a gauge-invariant local action without auxiliary fields. Therefore, consistency of (\ref{YM3}) is not automatic. Yet, rather surprisingly, the gauge covariant divergence of 
(\ref{YM3}) vanishes on-shell upon iterating the equation. 
This novel way of satisfying the consistency requirement is referred to as `third-way consistent' and reviewed in \cite{Bergshoeff:2015zga, Deger:2021ojb}. The mechanism was first discovered in the context of three-dimensional, higher curvature gravity theories \cite{Bergshoeff:2014pca} and other three-dimensional gravity examples were found later \cite{Ozkan:2018cxj, Afshar:2019npk}. Recently, third-way consistent $p$-form theories in arbitrary dimensions have been constructed
\cite{Broccoli:2021pvv} where a generalization  of (\ref{YM3}) with higher derivative terms is also given. 
Moreover, in \cite{Deger:2021fvv} the ${\cal N}=1$ off-shell supersymmetric extension of (\ref{YM3}) 
has been obtained so that bosonic and fermionic field equations are mapped to each other under supersymmetry.
The existence of a supersymmetric version of (\ref{YM3}) calls for a better understanding of this model. It was observed
in \cite{Broccoli:2021pvv} that the equations (\ref{YM3}) are actually dual to a principal chiral sigma model. 
In this paper we will make this correspondence more concrete and use this fact to construct the coupling of (\ref{YM3}) to
${\cal N}=1$ supergravity. After decoupling gravity, the model reduces to the supersymmetric system of \cite{Deger:2021fvv}.

The key to a dual formulation of (\ref{YM3}) is the on-shell duality between vector and scalar fields in three dimensions.
This is a standard feature for the dynamics of abelian vector fields satisfying the free Maxwell equations, but the duality can be extended
to non-abelian gauge groups and is central to the construction of gauged supergravities in three dimensions~\cite{Nicolai:2000sc}.
In particular, the topologically massive Yang-Mills Lagrangian (\ref{YMCS}) has a dual description 
as a gauged sigma model on the flat target space ${\mathfrak{g}}={\rm Lie}\,G$, upon introducing an additional set of vector fields $B_\mu$ gauging the shift symmetry on the scalar fields~\cite{Nicolai:2003bp}. Specifically, the dual Lagrangian is given by
\bea
\tilde{\cal L}_{\rm \tiny TMYM}&=& -\frac12\,{\rm tr}\left[{\cal D}_\mu \phi {\cal D}^\mu \phi \right]
+\varepsilon^{\mu\nu\rho}\, {\rm tr}\left[ B_\mu F_{\nu\rho} \right]
+2\,\mu\,\varepsilon^{\mu\nu\rho} \,{\rm tr}\left[ A_\mu{} \Big(
\partial_\nu A_{\rho} +\frac13\, \left[A_{\nu} ,A_{\rho}\right] \Big)  \right]
\;,\qquad
\label{LCSYM_dual}
\eea
with scalar fields $\phi$ in the adjoint representation of $\mathfrak{g}$, and their covariant derivatives defined as
\bea
{\cal D}_\mu \phi &=& \partial_\mu \phi +\left[ A_\mu{} , \phi \right] + B_\mu
\;.
\eea
In the dual formulation (\ref{LCSYM_dual}), all degrees of freedom are carried by the scalar fields $\phi$,
while the vector fields appear with a Chern-Simons rather than a Yang-Mills coupling. The full gauge algebra of (\ref{LCSYM_dual})
is given by $\mathfrak{g} \oright \mathfrak{n}$ where $\mathfrak{n}$ denotes a set of nilpotent generators
transforming in the adjoint representation of $\mathfrak{g}$.

In this note, we point out that a very similar construction can be given for the dual formulation of (\ref{YM3}). In this case, the scalar sigma model is built on the curved target space $G$, with gauging of the full $\mathfrak{g}\oplus\mathfrak{g}$ algebra of isometries. The Lagrangian is given by the gauged sigma model coupled to $G\times G$ Chern-Simons vector fields, see (\ref{LLL}) below. The rest of this note is organized as follows. In section~\ref{sec:dual}, we spell out the details of the dual formulation of (\ref{YM3}). In section~\ref{sec:coupling} we derive the coupling of (\ref{YM3}) to matter and to gravity directly in the dual formulation.
In section~\ref{sec:sugra} finally, we use the dual formulation together with the results of \cite{deWit:2003ja} to work out the coupling of (\ref{YM3}) to ${\cal N}=1$ supergravity.

\section{Dual formulation}
\label{sec:dual}

The dual formulation of (\ref{YM3}) is a gauged sigma model with Chern-Simons gauge fields.
Explicitly, we let the scalar fields of the model parametrize a ${\rm G}$-valued matrix $U$ with covariant derivatives
given by
\bea
{\cal D}_\mu U &=& \partial_\mu U +\beta\,B_\mu \,U -\alpha \,U A_\mu
\;,
\eea
i.e.\ the vector fields $A_\mu$ and $B_\mu$ gauge the commuting right and left group action on $U$
with coupling constants $\alpha$ and $\beta$, respectively. The left invariant currents
\bea
{\cal J}_\mu &=& U^{-1} {\cal D}_\mu U~\in~\mathfrak{g}
\;,
\label{J}
\eea
satisfy the integrability relations
\bea
{\cal D}_{[\mu} {\cal J}_{\nu]} + \frac12\left[{\cal J}_\mu,{\cal J}_\nu \right] &=& -\frac{\alpha}{2}\,F_{\mu\nu} +\frac{\beta}{2}\,U^{-1} H_{\mu\nu} U
\;,
\label{int}
\eea
with the field strengths given by
\bea
F_{\mu\nu} &=& 2\,\partial_{[\mu} A_{\nu]} + \alpha\,\left[A_\mu,A_\nu\right] \;,\nonumber\\
H_{\mu\nu} &=& 2\,\partial_{[\mu} B_{\nu]} + \beta\,\left[B_\mu,B_\nu\right] \;.
\eea
The  model is given by coupling the gauged sigma model with target space ${\rm G}$ built from (\ref{J}) 
to a ${\rm G}\times{\rm G}$ Chern-Simons term for the vector fields
\bea
{\cal L}_\sigma &=& -\frac12\,{\rm tr}\left[ {\cal J}_\mu {\cal J}^\mu \right]
- g_1\,
\varepsilon^{\mu\nu\rho} \,{\rm tr}\left[ A_\mu{} \Big(
\partial_\nu A_{\rho} +\frac{\alpha}3\, \left[A_{\nu} ,A_{\rho}\right] \Big)  \right]
\nonumber\\
&&{}
- g_2\,
\varepsilon^{\mu\nu\rho} \,{\rm tr}\left[ B_\mu{} \Big(
\partial_\nu B_{\rho} +\frac{\beta}3\, \left[B_{\nu} ,B_{\rho}\right] \Big)  \right]
\;,
\label{LLL}
\eea
with coupling constants $g_1$ and $g_2$, respectively. Variation of (\ref{LLL}) in particular yields the first-order field equations
\bea
{\cal J}_\mu &=& \frac{g_1}{\alpha}\,\varepsilon_{\mu\nu\rho}\,F^{\nu\rho}~=~
-\frac{g_2}{\beta}\,\varepsilon_{\mu\nu\rho}\,U^{-1} H^{\nu\rho} U
\;.
\label{JFH}
\eea
Plugging these relations back into the integrability relations (\ref{int}) yields
\bea
\varepsilon^{\mu\nu\rho}\,
  {\cal D}_{[\mu} {\tilde F}_{\nu]} + \frac{g_1}{\alpha}\ \varepsilon^{\mu\nu\rho}\, \left[{\tilde F}_\mu,{\tilde F}_\nu \right] 
  &=& -\frac12\left(\frac{\alpha^2}{g_1}+\frac{\beta^2}{g_2}\right) {\tilde F}^{\rho}
  \;.
  \label{YM3D}
\eea
Setting $\alpha=\beta=1$, these equations precisely reproduce the deformed Yang-Mills equations~(\ref{YM3})
with the translation of parameters
\bea
m&=&\frac{1}{2g_1}\;,\qquad
\mu-m ~=~\frac{1}{2g_2}
\;.
\label{mug}
\eea

Since both left and right group multiplication of $U$ are local symmetries, we
may fix part of the gauge symmetry by setting
\bea
U={\rm I}\;,
\eea
which breaks the gauge group down to the diagonal ${\rm G}_{\rm diag}\subset {\rm G}\times {\rm G}$.
With the translation of constants (\ref{mug}), the action (\ref{LLL}) then reduces to
\bea
{\cal L}_0 &=& -\frac12\,{\rm tr}\left[ (B_\mu-A_\mu) (B^\mu-A^\mu) \right]
- \frac{1}{2m}\,
\varepsilon^{\mu\nu\rho} \,{\rm tr}\left[ A_\mu{} \Big(
\partial_\nu A_{\rho} +\frac{1}3\, \left[A_{\nu} ,A_{\rho}\right] \Big)  \right]
\nonumber\\
&&{}
- \frac{1}{2(\mu-m)}\,
\varepsilon^{\mu\nu\rho} \,{\rm tr}\left[ B_\mu{} \Big(
\partial_\nu B_{\rho} +\frac{1}3\, \left[B_{\nu} ,B_{\rho}\right] \Big)  \right]
\;,
\label{LLL0}
\eea
which (up to a global factor) precisely reproduces the massive Chern-Simons 
action found in~\cite{Arvanitakis:2015oga}.
We note in passing that while the Lagrangian (\ref{LLL0}) appears degenerate at the 
special value $m=\mu$ ~\cite{Arvanitakis:2015oga}, in this case an action for the system
is straightforwardly obtained from the Lagrangian
(\ref{LLL}) upon setting $\beta=0=g_2$, such that the vector fields $B_\mu$
disappear from the action. Its gauge fixed version then reduces to the massive Chern-Simons Lagrangian
 \bea
{\cal L}_{0,\mu=m} &=& -\frac12\,{\rm tr}\left[ A_\mu A^\mu \right]
- \frac{1}{2m}\,
\varepsilon^{\mu\nu\rho} \,{\rm tr}\left[ A_\mu{} \Big(
\partial_\nu A_{\rho} +\frac{1}3\, \left[A_{\nu} ,A_{\rho}\right] \Big)  \right]
\;.
\label{LLL00}
\eea

Note that the dual formulation (\ref{LLL}) as well as its gauge fixed version (\ref{LLL0})
make manifest the duality symmetry of the system under simultaneous exchange of the vector fields 
$A_\mu$ and $B_\mu$ together with the parameters
\bea
g_1 \leftrightarrow g_2 \qquad
\Longleftrightarrow \qquad
m \leftrightarrow \mu - m
\;.
\label{dualmmu}
\eea
Since only the vector fields $A_\mu$ appear in the original deformed Yang-Mills equations (\ref{YM3}),
this duality can be employed in order to map solutions to (\ref{YM3}) with given values of $m$ and $\mu$
into new solutions to the same equations with different values of the parameters related by (\ref{dualmmu}).

Let us also note, that the equations (\ref{YM3}) admit a smooth limit to a deformation of
the Maxwell equations (i.e.\ to abelian field strengths $F_{\mu\nu}{}^A=2\,\partial_{[\mu}A_{\nu]}{}^A$) 
while keeping the structure constants $f_{BC}{}^A$ on the r.h.s.\ of (\ref{YM3}),
parametrizing the non-linear deformation of the topologically massive Maxwell equations as was considered in \cite{Broccoli:2021pvv}.
This limit is described by rescaling vector fields as
\bea
A_\mu{}^A &\rightarrow& m\,A_\mu{}^A
\;,
\label{scaleA}
\eea
dividing a resulting overall factor $m$ from (\ref{YM3}), and finally sending $m\rightarrow0$\,.
It is however not clear how this limit can be implemented directly on the level of the action (\ref{LLL}).

Finally, the limit of (\ref{LLL}) to the dual formulation of the
undeformed Yang-Mills theory (\ref{LCSYM_dual}) requires flattening of the sigma model target space,
such that in particular its isometry group contracts from ${\rm G}\times{\rm G}$ to ${\rm G}\ltimes{\rm N}$\,.

\section{Coupling to matter and gravity}
\label{sec:coupling}

The coupling of the deformed Yang-Mills system (\ref{YM3}) to gravity and lower-spin matter has been 
determined in~\cite{Arvanitakis:2015oga} by a careful analysis of the consistency conditions of the coupled system.
Within the dual formulation (\ref{LLL}), this coupling is straightforward. 
Simply adding a matter Lagrangian ${\cal L}_{\rm mat}$ to (\ref{LLL})
\bea
{\cal L} &=&{\cal L} _\sigma+{\cal L}_{\rm mat}
\;,
\label{LLmat}
\eea
the first-order field equations (\ref{JFH}) change into
\bea
{\cal J}_\mu &=& \frac{g_1}{\alpha}\,\varepsilon_{\mu\nu\rho}\,F^{\nu\rho} +j_\mu\;,\qquad
j_\mu= \frac1{\alpha}\, \frac{\partial{\cal L}_{\rm mat}}{\partial A^\mu}
\;.
\label{JFHmat}
\eea
For simplicity, we assume that the matter is only charged under the right factor
of the gauge group ${\rm G}\times{\rm G}$, i.e.\ that ${\cal L}_{\rm mat}$  does not depend on $B_\mu$.
After plugging these equations into the integrability relations (\ref{int}),
the equations (\ref{YM3D}) are modified according to
\bea
\varepsilon^{\mu\nu\rho}\,
  {\cal D}_{[\mu} {\tilde F}_{\nu]} + \frac{g_1}{\alpha}\ \varepsilon^{\mu\nu\rho}\, \left[{\tilde F}_\mu,{\tilde F}_\nu \right] 
  +\frac12\left(\frac{\alpha^2}{g_1}+\frac{\beta^2}{g_2}\right) {\tilde F}^{\rho}
  &=& 
- \frac{\alpha}{2g_1}\, \varepsilon^{\mu\nu\rho}\, {\cal D}_{\mu} j_{\nu} 
 - \frac{\alpha}{4g_1} \, \varepsilon^{\mu\nu\rho} \left[j_\mu,j_\nu \right] 
 \nonumber\\
 &&{}
 -\frac{\alpha\,\beta^2}{4\,g_1g_2}\, j^{\rho}
  -  \varepsilon^{\mu\nu\rho}\, \big[{\tilde F}_\mu,j_\nu \big] 
  \;,
  \label{DFJJ}
\eea
where the r.h.s.\ directly reproduces (after setting $\alpha=\beta=1$,
rescaling $j_\mu \rightarrow -4\,g_1g_2\,j_\mu$,
and translating parameters by (\ref{mug})) the matter
coupling constructed in~\cite{Arvanitakis:2015oga} but with the opposite sign of the second term on the r.h.s. which agrees with \cite{Broccoli:2021pvv}.

Similarly, coupling of the system (\ref{LLL}), (\ref{LLmat}) to three-dimensional gravity is straightforward. 
The full energy-momentum tensor $T_{\mu\nu}$ is given by
\bea
T_{\mu\nu} &=& 
-\frac1{2m^2}\,{\rm tr}\big[{\tilde F}_\mu {\tilde F}_\nu\big]
+\frac1{4m^2}\,g_{\mu\nu}\,{\rm tr}\big[{\tilde F}_\rho {\tilde F}^\rho\big]
-\frac1{m}{\rm tr}\big[{\tilde F}_\mu {j}_\nu\big]
+\frac1{2m}\,g_{\mu\nu}\,{\rm tr}\big[{\tilde F}_\rho {j}^\rho \big]
\nonumber\\
&&{}
-\frac12\,{\rm tr}\big[j_\mu j_\nu\big]
+\frac14\,g_{\mu\nu}\,{\rm tr}\big[j_\rho j^\rho \big]
+\sqrt{|g|}^{-1}\,\frac{\partial {\cal L}_{\rm mat}}{\partial g^{\mu\nu}}
\;,
\label{TT}
\eea
with the first six terms derived from $\sqrt{|g|}^{-1}\frac{\partial {\cal L}_{\sigma}}{\partial g^{\mu\nu}}$ 
after using (\ref{JFHmat}).
In absence of ${\cal L}_{\rm mat}$, i.e.\ for $j_\mu=0$, this precisely
reproduces the energy-momentum tensor given in~\cite{Arvanitakis:2015oga}. 
The full expression (\ref{TT}) directly extends this
result to the presence of additional matter.

\section{Coupling to ${\cal N}=1$ supergravity}
\label{sec:sugra}

The ${\cal N}=1$ supersymmetrization of the system (\ref{YM3}) was 
constructed in~\cite{Deger:2021fvv} on the level of the field equations.
In terms of the dual formulation~(\ref{LLL}), we can trace this back to the 
known structures of a supersymmetric sigma model with target space ${\rm G}$~\cite{Alvarez-Gaume:1981exv}.
As a new application, we will employ the dual formulation~(\ref{LLL}) in order to
work out the coupling of the system (\ref{YM3}) to ${\cal N}=1$ supergravity. 
The coupling of a general gauged sigma-model to three-dimensional supergravity has been
constructed in~\cite{deWit:2003ja}. In particular, any Riemannian target space
allows for a coupling to ${\cal N}=1$ supergravity. The fermionic field content in this case
is given by $n(={\rm dim\,G})$ spin-1/2 fermions $\chi^i$ together with a gravitino $\psi_\mu$.

To translate to the notation from~\cite{deWit:2003ja}, we parametrize the group manifold ${\rm G}$
by coordinates $\phi^i$ and define the left and right invariant vector fields $L_A$, $R_A$ satisfying the 
differential relations
\bea
\nabla_i L_{A\,j} = -\frac12\,{f}_{A}{}^{BC}\,L_{B\,i} L_{C\,j} \;,\qquad
\nabla_i R_{A\,j} = \frac12\,{f}_{A}{}^{BC}\,R_{B\,i} R_{C\,j} 
\;.
\eea
The metric on the group manifold can be expressed as
\bea
g_{ij} &=& L_{A\,i} L^A_j ~=~ R_{A\,i} R^A_j
\;.
\eea
The currents of the gauged sigma model (\ref{J}) are then given by
\bea
{\cal J}_\mu{}^A &=& L^A_{i }\,{\cal D}_\mu\phi^i ~=~ L^A_{i }\, \partial_\mu \phi^i 
-\alpha \,A^A_\mu 
+\beta \,B^B_\mu \, ( L^A_{i} R_B^{i}) 
\;.
\eea
On the fermionic side, we define the fermions $\chi^A=L_i^A\chi^i$ transforming under the 
right action of ${\rm G}$, with covariant derivatives
\bea
{\cal D}_\mu \chi^A &=&
 \left( \partial_\mu +\tfrac12 \omega_\mu{}^a \gamma_a \right)\chi^{A}  + \alpha {f}_{BC}{}^{A}\, A_\mu^B \chi^C
 \;.
\eea
In terms of these objects, we can obtain the coupling of the gauged sigma model (\ref{LLL})  to
${\cal N}=1$ supergravity from the general result of~\cite{deWit:2003ja} 
(after a suitable rescaling of the vector fields)
as
\bea
\sqrt{g}^{-1}{\cal L} &=& -\ft12\,\epsilon^{\mu\nu\rho}\left(
e_\mu{}^a\,R_{\nu\rho a} +
\bar\psi{}_\mu {\cal D}_\nu\psi_\rho  \right)
-4\,C^2
- g_1\,
\epsilon^{\mu\nu\rho} \, A_\mu{}^A \Big(
\partial_\nu A_{\rho}{}^A +\tfrac{\alpha}3\, f_{BC}{}^A A_{\nu}{}^B A_{\rho}{}^C \Big) 
\nonumber\\
&&{}
- g_2\,
\epsilon^{\mu\nu\rho} \, B_\mu{}^A \Big(
\partial_\nu B_{\rho}{}^A +\tfrac{\beta}3\, f_{BC}{}^A B_{\nu}{}^B B_{\rho}{}^C \Big) 
-\ft12  {\cal J}_{\mu}{}^{A} {\cal J}^{\mu A}
-\ft12  \, \bar\chi{}^{A} \gamma^\mu {\cal D}_\mu \chi^A 
\nonumber\\
&&{}
+\ft14  {\cal J}_{\mu}{}^{A}\,  {f}^A{}_{BC}  \bar\chi{}^{B} \gamma^\mu  \chi^{C}  
+ \ft12  {\cal J}_{\nu}{}^{A}\,  \bar\chi{}^{A}\gamma^\mu\gamma^\nu
\psi_\mu
+
\ft12 \,C\,\bar\psi{}_\mu\,\gamma^{\mu\nu}\,\psi_\nu\, 
\nonumber\\
&&{}
-
\tfrac14 \left(\tfrac{\alpha^2}{g_1}+\tfrac{\beta^2}{g_2}-2C \right) \bar\chi^{A}\chi^{A}
+  \ft1{16} \, (\bar\chi{}^{A} \chi^{A})^2 
- \ft18 \,   (\bar\chi{}^{A}\gamma^\mu\gamma^\nu
\psi_\mu)\,(\bar\psi_\nu \chi^A)
\nonumber\\
&&{}
+\ft1{96}  \,{f}_{AB}{}^E {f}_{CDE}\; \bar\chi{}^{A} \gamma_\mu
\chi^{B} \, \bar\chi{}^{C} \gamma^\mu \chi^{D}  
\;,
\label{LN1}
\eea
with a cosmological constant $C$. The supersymmetry variations are given by
\begin{eqnarray}
\delta e_\mu{}^a &=& \ft12\,\bar \epsilon{}\gamma^a\,\psi_\mu 
\;,\qquad \delta \phi^i ~=~ \ft12\,\bar \epsilon\,\chi^{i} 
\;,\nonumber\\
\delta A_\mu{}^A &=& \tfrac{\alpha}{4g_1}\,\bar \chi^{A}\gamma_\mu\epsilon
\;,\qquad
\delta B_\mu{}^A ~=~ -\tfrac{\beta}{4g_2}\, ( R^A_{i} L^i_B) \, \bar \chi^{B}\gamma_\mu\epsilon
\;,\nonumber\\
\delta \psi_\mu &=& \left( \partial_\mu +\tfrac12 \omega_\mu{}^a \gamma_a \right) \epsilon   - \ft18 \,\bar \chi{}^{A}
\gamma^\nu \chi^{A}\, \gamma_{\mu\nu} \,\epsilon 
+C \gamma_\mu\,\epsilon 
 \;,\nonumber\\
\delta \chi^{A} &=& \tfrac12\, {{\cal J}}_\mu{}^A \,
\gamma^\mu   \epsilon
-
\tfrac14\, (\bar \psi{}_\mu \chi^{A} ) \,
\gamma^\mu  \epsilon
-
\tfrac14\,{f}_{BC}{}^A\, ( \bar \epsilon\,\chi^{B})\, \chi^{C} 
\;,
\label{susytraf}
\end{eqnarray}
and leave the Lagrangian (\ref{LN1}) invariant. The supersymmetric extension of the 
Yang-Mills system (\ref{YM3}) is then obtained in full analogy to (\ref{YM3D}), and (\ref{DFJJ}). 
Variation of (\ref{LN1}) w.r.t.\ the vector fields, 
yields the first-order duality equations
\bea
 {\cal J}^{\mu\,A}
&=&
\frac{g_1}{\alpha}\, \epsilon^{\mu\nu\rho} \,F_{\nu\rho}{}^A 
+\frac{1}{2}\,   \bar \chi{}^{A}\gamma^\nu\gamma^\mu\,\psi_\nu
- \frac{1}{4}\, {f}^A{}_{BC} \, \bar \chi{}^{B} \gamma^\mu  \chi^{C} 
\nonumber\\
&=&
-\frac{g_2}{\beta}\,  \epsilon^{\mu\nu\rho}\,H_{\nu\rho}{}^{B} \,(R_B^i L^A_i )
+\frac{1}{2}\,   \bar \chi{}^{A}\gamma^\nu\gamma^\mu\,\psi_\nu
+\frac{1}{4}\, {f}^A{}_{BC} \, \bar \chi{}^{B} \gamma^\mu  \chi^{C} 
\label{JAB}
\;.
\eea
Upon plugging these equations into the integrability relations~(\ref{int})
of the scalar currents, 
and translating the coupling constants via (\ref{mug}) together with $\alpha=\beta=1$,
we finally obtain the desired supersymmetric extension of (\ref{YM3})
\bea
\epsilon^{\mu\nu\rho}\,
  {\cal D}_{[\mu} {\tilde F}_{\nu]}{}^A 
 +\mu\, {\tilde F}^{\rho\,A}
   &=& 
   - \frac{1}{2m}\ \epsilon^{\mu\nu\rho}\, f_{BC}{}^A \, {\tilde F}_\mu{}^B {\tilde F}_\nu{}^C 
  -  \epsilon^{\mu\nu\rho}\, f_{BC}{}^A\,{\tilde F}_\mu\,^B j^{(+)C}_\nu 
- m\,\epsilon^{\mu\nu\rho}\, {\cal D}_{\mu} j^{(+)A}_{\nu} 
 \nonumber\\
 &&{}
 +m(m-\mu)\left( j^{(+)\rho A} -  j^{(-)\rho A}  \right)
 - \frac{m}{2} \, \epsilon^{\mu\nu\rho}\,  f_{BC}{}^A\,j^{(+)B}_\mu  j^{(+)C}_\nu
  \;,
  \label{YM3N1}
\eea
with matter currents bilinear in the fermions
\bea
{j}_{\mu}^{(\pm)\,A}
&\equiv&
 \frac12\,  \bar \chi{}^{A}\gamma^\nu\gamma_\mu\,\psi_\nu
\mp \frac14\,  {f}_{BC}{}^A \, \bar \chi{}^{B} \gamma_\mu  \chi^{C} 
\;.
\eea
The appearance of a second current ${j}_{\mu}^{(-)}$ as opposed to the general form of (\ref{DFJJ})
is due to the appearance of the vectors $B_\mu{}^A$ via the fermion couplings to ${\cal J}_\mu{}^A$ in the matter couplings of (\ref{LN1}).

In order to complete the supersymmetric system, it remains to compute the remaining equations of motion from (\ref{LN1})
and replace all the appearing scalar currents ${\cal J}_\mu{}^A$ and field strengths $H_{\mu\nu}{}^A$ by means of the duality equations (\ref{JAB}).
For the gravitino and the spin 1/2 field equations, this yields after some straightforward computation (and notably some Fierzing) the deformed Rarita-Schwinger equation
\bea
\epsilon^{\mu\nu\rho}
  {\cal D}_\nu\psi_\rho 
 &=& 
 \ft1{2m}   \gamma^\nu \gamma^\mu  \chi{}^{A}\,\tilde{F}_{\nu}{}^{A}
+ C\,\gamma^{\mu\nu}\,\psi_\nu\, 
- \ft1{4}\, {f}_{BC}{}^A  \chi{}^{A}\, \,( \bar \chi{}^{B} \gamma^\mu  \chi^{C})
- \ft14  \, (\bar \chi{}^{A} \chi^A) \, \psi^\mu
\;,\;\;\;
\label{eom_gravitino}
\eea
as well as the deformed Dirac equation
\bea
\gamma^\mu {\cal D}_\mu \chi^A 
  &=& 
\ft1{2m}  \gamma^\mu  \chi^{B} \tilde{F}_{\mu}{}^{C} \,{f}_{BC}{}^A 
+ \ft1{2m}   \gamma^\mu\gamma^\nu \psi_\mu\,\tilde{F}_{\nu}{}^{A}
+  \left(C -\mu \right) \chi^{A}
+\ft1{4}  \psi_\nu \,{f}_{BC}{}^A  (\bar \chi{}^{B} \gamma^\nu \chi^C)
\nonumber\\
&&{}
- \ft14    \chi^A (\bar \psi_\mu \psi^\mu)
-\ft1{12}  \gamma^\mu  \chi^{B} (\bar \chi{}^{D} \gamma_\mu  \chi^{E} ) \,{f}_{BC}{}^A  {f}_{DE}{}^C 
+  \ft1{4} \,\chi^A\, (\bar \chi{}_B \chi^{B}) 
\;.
\label{eom_fermion}
\eea
Finally, the Einstein field equations are of the standard form with the energy-momentum tensor given by (\ref{TT})
and again all scalar currents eliminated by (\ref{JAB}).
By construction, the system of (\ref{YM3N1}), (\ref{eom_gravitino}), (\ref{eom_fermion}), and the 
Einstein equations yields a supersymmetric system, with the equations transforming into each other 
under the transformations (\ref{susytraf}) of the remaining fields
\begin{eqnarray}
\delta e_\mu{}^a &=& \tfrac12\,\bar \epsilon{}\gamma^a\,\psi_\mu 
\;,\nonumber\\
\delta A_\mu{}^A &=& \tfrac{m}2 \bar \chi^{A}\gamma_\mu\epsilon
\;,\nonumber\\
\delta \psi_\mu &=& \left( \partial_\mu +\tfrac12 \omega_\mu{}^a \gamma_a \right) \epsilon    - \tfrac18 \,\bar \chi{}^{A}
\gamma^\nu \chi^{A}\, \gamma_{\mu\nu} \,\epsilon 
+C \gamma_\mu\,\epsilon 
 \;,\nonumber\\
\delta \chi^{A} 
&=&
 \tfrac1{2m}\, {{\tilde F}}_\mu{}^A  \,\gamma^\mu   \epsilon
+ \tfrac14\,(\bar \chi^A \gamma^{\mu\nu}  \psi_\mu)   \,\gamma_\nu   \epsilon
\;.
\label{susytraf_final}
\end{eqnarray}
A direct construction of this system starting from the supersymmetrization of (\ref{YM3N1})
would have represented a formidable technical challenge with an uncertain outcome. In the dual picture,
the entire system is straightforwardly obtained by variation of the supersymmetric action (\ref{LN1}).

As a consistency check, we may study the rigid limit of the model. In order to decouple gravity, we first set $\psi_{\mu}=C=0$, 
and then scale fields and structure constants with a constant $k$ as $\chi^{A} \rightarrow k\chi^{A}, A_{\mu}{}^A \rightarrow kA_{\mu}{}^A, f_{BC}{}^A \rightarrow f_{BC}{}^A/k$. After 
cancelling the overall $k$ and taking the limit $k \rightarrow 0$ together with a few redefinitions,\footnote{Explicitly, $m \rightarrow m/4, \, \mu \rightarrow 2\mu, \, {j}_{\mu}^{(+)\,A} \rightarrow -2j_{\mu}^{\,A}/m$.}
one arrives precisely at the field equations of ${\cal N}=1$ off-shell supersymmetric massive Yang-Mills theory constructed in \cite{Deger:2021fvv}. This limit can also be taken at the action \eqref{LN1} by setting $g_{\mu\nu}=\eta_{\mu\nu}$ and making additional scalings $\phi^i \rightarrow k\phi^i$, $ B_{\mu}{}^A \rightarrow kB_{\mu}{}^A$.

Let us finally note that the field equations (\ref{YM3N1}), (\ref{eom_gravitino}), (\ref{eom_fermion}), as well as the
supersymmetry transformations (\ref{susytraf_final}) are compatible with the limit $m\rightarrow0$ after rescaling
the vector fields as (\ref{scaleA}). This is the limit to abelian field strengths $F_{\mu\nu}{}^A=2\,\partial_{[\mu}A_{\nu]}{}^A$
while keeping all the remaining structure constants $f_{BC}{}^A$ in the various couplings \cite{Broccoli:2021pvv}.

\section{Conclusions}

In this note, we have shown that the dual formulation~(\ref{LLL}) of the deformed Yang-Mills equations (\ref{YM3}) 
provides a natural explanation
for many features of the model and in particular allows for a straightforward coupling of the model to ${\cal N}=1$ supergravity.
The dual formulation also shows that the construction cannot be extended to higher supersymmetry ${\cal N}>1$, 
since for a sigma model higher supersymmetry requires further constraints on the scalar target space~\cite{deWit:1992up} 
which are not satisfied by the real compact Lie groups $G$ that appear here. 
In particular, an extension to ${\cal N}=2$ would require $G$ to be a
K\"ahler manifold, thus the gauge group to be complex \cite{complex}.

Among further applications, it would be interesting to work out the analogous dual models to the higher dimensional third way consistent $p$-form theories of \cite{Broccoli:2021pvv}.  Higher
derivative extensions of \eqref{YM3} were constructed in \cite{Broccoli:2021pvv} and finding supersymmetric versions of these would also be desirable. 

Another interesting extension of the present construction is its analogue in the context
of third way consistent gravitational theories, such as minimal massive gravity\cite{Bergshoeff:2014pca}, and extensions  thereof \cite{Ozkan:2018cxj, Afshar:2019npk}. We will come back to this in \cite{DGRSxx}.

\section*{Acknowledgements}
We would like to thank Marc Geiller and Jan Rosseel for discussions and Craig van Coevering for pointing out reference \cite{complex} to us. NSD is grateful to ENS de Lyon, Laboratoire de Physique for hospitality 
where this work was initiated and Research
Fellowship Program of Embassy of France in Turkey for financial support during this visit. NSD would also like to thank ESI, Vienna for hospitality and financial support
where this work was completed.

\providecommand{\href}[2]{#2}\begingroup\raggedright\endgroup


\end{document}